\documentclass[aps,prx,amsmath,amssymb,amsfonts,superscriptaddress,notitlepage,reprint,longbibliography,floatfix
]{revtex4-2}
\usepackage{graphicx}
\usepackage{dcolumn}
\usepackage{bm}
\usepackage{amssymb}
\usepackage{amsmath}
\usepackage[svgnames]{xcolor}
\usepackage[colorlinks=true,
            linkcolor=red,
            urlcolor=blue,
            citecolor=DarkGreen]{hyperref}
\usepackage[english]{babel}
\usepackage{braket}
\usepackage{microtype}
\usepackage{bbm}

\begin{document}
\title{Realization of a Universal Quantum Gate Set for Itinerant Microwave Photons}

\author{Kevin Reuer}\email{kevin.reuer@phys.ethz.ch}
\author{Jean-Claude~Besse}
\author{Lucien Wernli}
\author{Paul Magnard}
\author{Philipp Kurpiers}
\author{Graham J. Norris}
\affiliation{Department of Physics, ETH Zurich, CH-8093 Zurich, Switzerland}
\author{Andreas Wallraff}
\affiliation{Department of Physics, ETH Zurich, CH-8093 Zurich, Switzerland}
\affiliation{Quantum Center, ETH Zurich, CH-8093 Zurich, Switzerland}
\author{Christopher Eichler}\email{eichlerc@phys.ethz.ch}
\affiliation{Department of Physics, ETH Zurich, CH-8093 Zurich, Switzerland}

\date{\today}

\begin{abstract}
Deterministic photon-photon gates enable the controlled generation of entanglement between mobile carriers of quantum information. Such gates have thus far been exclusively realized in the optical domain and by relying on post-selection. Here, we present a non-post-selected, deterministic, photon-photon gate in the microwave frequency range realized using superconducting circuits. We emit photonic qubits from a source chip and route those qubits to a gate chip with which we realize a universal gate set by combining controlled absorption and re-emission with single-qubit gates and qubit-photon controlled-phase gates. We measure quantum process fidelities of $75\,\%$ for single- and of $57\,\%$ for two-qubit gates, limited mainly by radiation loss and decoherence. This universal gate set has a wide range of potential applications in superconducting quantum networks.
\end{abstract}

\maketitle
Photons are ideal carriers of quantum information because they do not interact with each other when propagating through linear media. Such interactions, on the other hand, are essential for the implementation of photon-photon gates \cite{Kok2007}, which can be used within quantum networks \cite{Kimble2008,Xiang2017} or in distributed quantum computing \cite{Cirac1999,Pichler2016,Nickerson2014,Xiang2017}. Photon-photon gates enable, for example, the processing of quantum information while photonic qubits are traveling between local quantum processing units, potentially allowing for simpler network structures \cite{Slussarenko2019}. All-photonic quantum computing paradigms, such as linear optics \cite{Knill2001,Kok2007,Pan2012,Slussarenko2019}, continuous-variable \cite{Menicucci2006, Arrazola2021} and photonic one-way \cite{Walther2005,Larsen2019,Asavanant2019} quantum computing also rely on photon-photon gates.


Probabilistic photon-photon gates are based on beam-splitter-induced interference effects and projective measurements \cite{Knill2001, Kok2007}. In particular, the bosonic symmetry relation \cite{Kok2007} gives rise to the Hong-Ou-Mandel effect \cite{Hong1987}, providing a resource for entanglement generation. Together with the non-linearity of photon detection, interference effects enable (probabilistic) Bell measurements \cite{Weinfurter1994, Kok2007} as used in quantum repeater protocols \cite{Briegel1998,Jiang2009a,Munro2012, Azuma2015}.  By additionally using ancilla measurements and classical feed forward, probabilistic photon-photon gates have been implemented in optical setups \cite{OBrien2003, Zhao2005, Lu2019a, Lo2020a, Zeuner2018, Carolan2015} and are now commonly used in linear optics quantum computing \cite{Carolan2015,Carolan2020,Zhong2020a}.

Alternatively, interactions of photons with non-linear media enable deterministic photon-photon gates \cite{Duan2004}. Such gates have recently been implemented in the optical domain, either based on photons sequentially reflecting off of an atom in a cavity \cite{Hacker2016b} or based on the transmission through an ensemble of Rydberg atoms after absorption of a control photon \cite{Tiarks2019}. Albeit both implementations \cite{Hacker2016b,Tiarks2019} are deterministic, they use post-selection to compensate for losses in the experimental setup.

Here, we exploit the controllable coupling between superconducting qubits and itinerant microwave fields to implement a deterministic, non-post-selected, photon-photon controlled-phase (CPHASE) gate. To this aim, we absorb a first photonic qubit using a stationary gate qubit, reflect off a second photonic qubit of the gate qubit to perform the entangling gate, and re-emit the first one. Additionally, we demonstrate universal single-qubit gates acting on the photonic qubits. Since we encode photonic qubits in the single-rail basis as single-photon microwave fields, single-qubit gates do not preserve the photon number and can thus not be realized with linear optics elements \cite{Kok2007}.

\begin{figure}[t]
\includegraphics[width=\columnwidth]{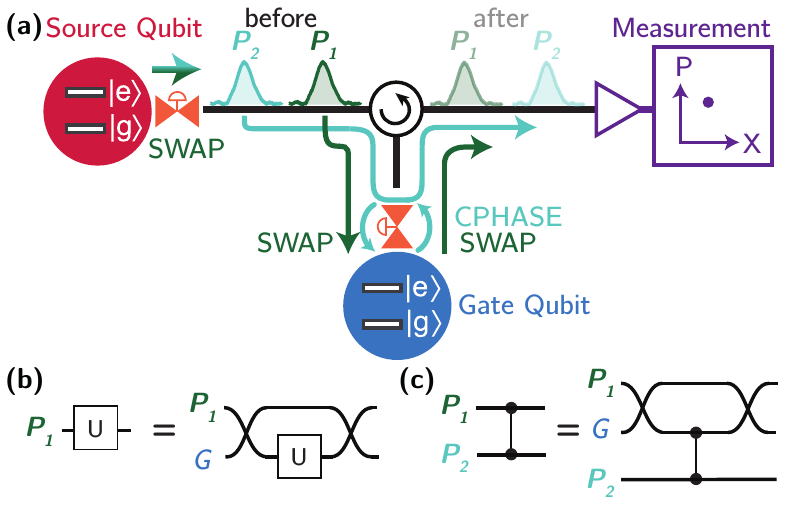}
\caption{\label{fig:Principle} (a) Schematic representation of the concept of the experiment. We generate photonic qubits $P_1$ (dark green) and $P_2$ (turquoise) from the source qubit (red) via swap gates (SWAP) mediated by a tunable coupling depicted as a control valve (orange). The photonic qubits propagate through a circulator to the gate qubit (blue), performing a swap gate (SWAP) or a controlled-phase gate (CPHASE). We linearly amplify the output field and measure the photonic qubits using heterodyne detection (purple).
(b,c) Equivalent quantum circuits implemented to realize (b) an arbitrary single-qubit gate $U$ and (c) a photon-photon controlled-phase (CPHASE) gate.
}
\end{figure}

 
We use a source qubit $S$ and a gate qubit $G$, both tunably coupled to the microwave field in a waveguide, as non-linear systems to generate photonic qubits and to temporarily store and act on qubit states, respectively, see schematic in Fig.~\ref{fig:Principle}. We first generate a train of itinerant photonic qubits $P_i$ in the single-rail basis by performing swap gates between the source qubit $S$ and the itinerant field carrying the photonic qubits $P_i$ \cite{Besse2020a}, as shown in Fig.~\ref{fig:Principle}(a). The photonic qubits $P_i$ propagate through a circulator towards the gate qubit $G$ [Fig.~\ref{fig:Principle}(a)], which implements both single- and two-qubit gates. To realize single-qubit gates acting on $P_1$, we first absorb $P_1$ by the gate qubit $G$ using the same swap gate as used for photon emission. We then apply a single-qubit unitary $U$ on the gate qubit $G$ and finally perform a second swap to re-emit the photonic qubit $P_1$, see equivalent quantum circuit in Fig.~\ref{fig:Principle}(b). For the controlled-phase gate [Fig.~\ref{fig:Principle}(c)] acting on two photonic qubits $P_1$ and $P_2$, we first absorb the control qubit $P_1$ and then reflect the target qubit $P_2$ off of the gate qubit $G$ resulting in a controlled-phase gate between $P_2$ and $G$ \cite{Besse2017,Kono2018}. Finally, we re-emit the control qubit $P_1$. After interacting with the gate qubit $G$, the photonic qubits $P_i$ propagate through the circulator to the output port at which we measure the field quadratures $X$ and $P$ of the photonic qubits $P_i$ using linear amplification and heterodyne detection to tomographically reconstruct quantum states and processes \cite{Besse2020a, Eichler2012}.

\begin{figure}[t]
\includegraphics[width=\columnwidth]{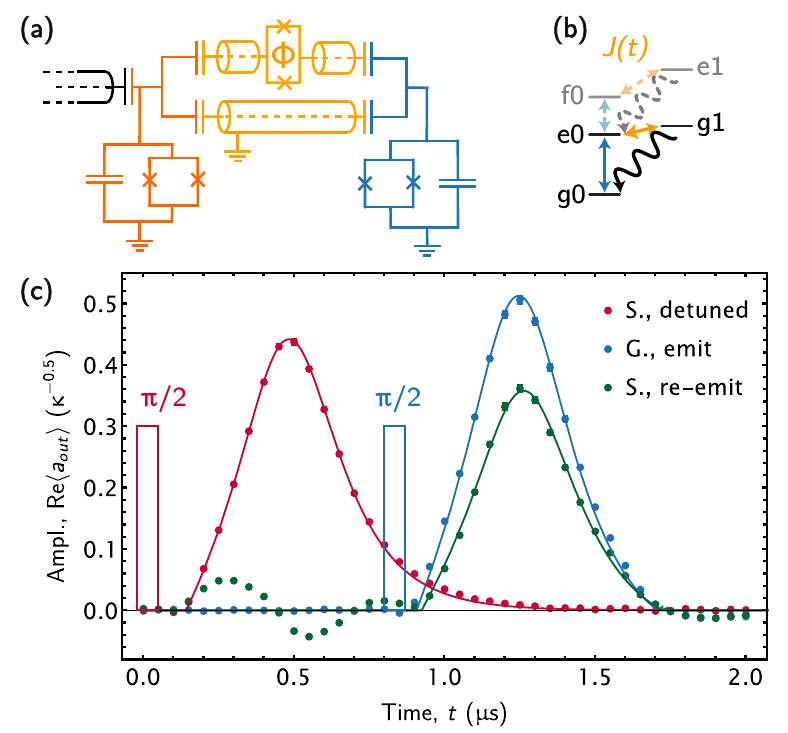}
\caption{\label{fig:experimental_setup}
(a) Electrical circuit diagram of the gate qubit (blue) interacting with fields in the transmission line (black) through a tunable coupler (orange) and a converter mode (dark orange, realized as a transmon qubit). (b) Energy level diagram indicating transitions induced by single-qubit gates (blue), parametric modulation of the flux $\Phi$ threading the SQUID loop embedded in the coupler (orange), and photon emission via spontaneous decay (black). (c) Measured (dots) and simulated (solid lines) real part of the envelope of the field amplitude $\text{Re}\langle a_{\text{out}} \rangle$ \emph{vs.}~time $t$ for superposition states $(\ket{0}+\ket{1})/\sqrt{2}$ emitted from the source qubit with the converter mode of the gate device being detuned from the field (red), when absorbing and re-emitting the field at the gate qubit (green), and when directly preparing and emitting the state at the gate qubit (blue). A red (blue) wireframe shows the timing of qubit pulses with a rotation angle of $\pi/2$ applied to the source (gate) qubit prior to emission. Imaginary parts, not shown, are below $0.07/\sqrt{\kappa}$ in magnitude.
}
\end{figure}

We experimentally realize the described scheme for single- and two-photon gates by using two copies of the superconducting chip presented in Ref.~\cite{Besse2020a} (optical micrograph in App.~\ref{SM:setup}), one for the source qubit $S$ and one for the gate qubit $G$, see Fig.~\ref{fig:experimental_setup}(a) for an equivalent electrical circuit diagram. The source qubit $S$ (gate qubit $G$) is realized as a flux-tunable transmon qubit tuned to its maximum transition frequency at $\omega_{S,ge}/2 \pi=5.925$~GHz  ($\omega_{G,ge}/2 \pi=5.771$~GHz), coupled to the transmission line via a tunable coupler at $\omega_{S,c}/2 \pi=3.2$~GHz ($\omega_{G,c}/2 \pi=4.6$~GHz) and a converter mode also realized as a transmon, biased to $\omega_{01}/2 \pi=5.998$~GHz. We perform local unitaries by applying microwave pulses via dedicated charge lines. For performing the swap gate, we activate the interaction Hamiltonian $H_{\text{SWAP}}=\hbar J (\ket{e0}\bra{g1}+ \ket{g1} \bra{e0})$, where $\ket{g}$ and $\ket{e}$ denote the ground and first excited state of the source qubit (gate qubit) and $\ket{0}$ and $\ket{1}$ denote the ground and first excited state of the converter mode, which couples with constant rate $\kappa_S/2 \pi=1.8$~MHz ($\kappa_G/2 \pi=2.1$~MHz) to a transmission line into which the propagating fields are emitted. Thus, the converter mode frequency sets the carrier frequency $\omega_{01}/2 \pi=5.998$~GHz of the generated photonic mode. As shown in Fig.~\ref{fig:experimental_setup}(b), we realize the interaction Hamiltonian $H_{\text{SWAP}}$ by parametrically driving the transition $\ket{e0} \leftrightarrow \ket{g1}$ via a magnetic flux pulse $\Phi(t)$ applied to the superconducting quantum interference device (SQUID) loop of the coupler with a carrier frequency matching the level detuning $\Delta_S=\omega_{01}/2 \pi-\omega_{S,ge}/2 \pi=73$~MHz ($\Delta_G=\omega_{01}/2 \pi-\omega_{G,ge}/2 \pi=227$~MHz). To optimize the absorption efficiency, we shape the temporal profile of the photonic mode $\xi(t) \propto \text{sech}(\Gamma t/2)$ to be time-symmetric with bandwidth $\Gamma$ \cite{Kurpiers2018,Pechal2014,Morin2019}, by controlling the interaction strength $J$ (see App.~\ref{SM:PS} for details). We further optimize the absorption process by tuning the converter modes of the source and gate device into resonance with each other (see App.~\ref{SM:Calib} for details).

To demonstrate our ability to emit and absorb photonic modes with time-symmetric profile, we measure the average field amplitude \emph{vs.}~time of photonic qubits prepared in the state $\ket{+}=(\ket{0}+\ket{1})/\sqrt{2}$ in three different experimental settings [Fig.~\ref{fig:experimental_setup}(c)]. First, we verify the emission of shaped photonic modes by emitting a $\ket{+}$ state from the source qubit $S$ with the converter mode of the gate chip detuned, such that the emitted photonic modes do not interact with the gate chip. The measured average field amplitude (red dots) is in good agreement with the temporal profile extracted from a two-level Rabi model using the calibrated parameters (red line, see App.~\ref{SM:PS} for details). Second, we demonstrate that we can absorb and re-emit shaped photonic modes by emitting a photonic qubit in the $\ket{+}$ state from the source qubit, and absorbing and re-emitting the photonic qubit at the gate qubit. The measured amplitude (green dots) is also in good agreement with the simulated one (green line), but is $19\,\%$ smaller in amplitude [compared to the photonic qubits emitted from the source qubit with the gate qubit converter mode detuned (red)], mainly due to decoherence of the gate qubit (inferred from a master equation simulation, see App.~\ref{SM:MES} for details). For comparison and to quantify the photon loss during propagation from the source to the gate device and during the absorption and re-emission process, we also emit $\ket{+}$ states directly prepared at the gate qubit (blue dots). We find a ratio between the measured amplitudes of photonic qubits emitted from the source (red, with gate qubit converter mode detuned) and gate qubit (blue) of about $86\,\%$ which is close to the transmission efficiency $\sqrt{\eta_{\text{loss}}} = \sqrt{1-0.25}=87\,\%$, determined in an independent calibration measurement (see App.~\ref{SM:setup} for details) and is mainly due to the loss in the circulator. From the measured photon field amplitudes, we estimate an absorption efficiency $\eta_{\text{abs}}$ of $97\,\%$, mainly limited by the truncation of the flux pulse generating the photonic mode $\xi(t) \propto \text{sech}(\Gamma t/2)$ at $\pm 4.6 / \kappa_S$ (see App.~\ref{SM:PS} for details).
\begin{figure}[t]
\includegraphics[width=\columnwidth]{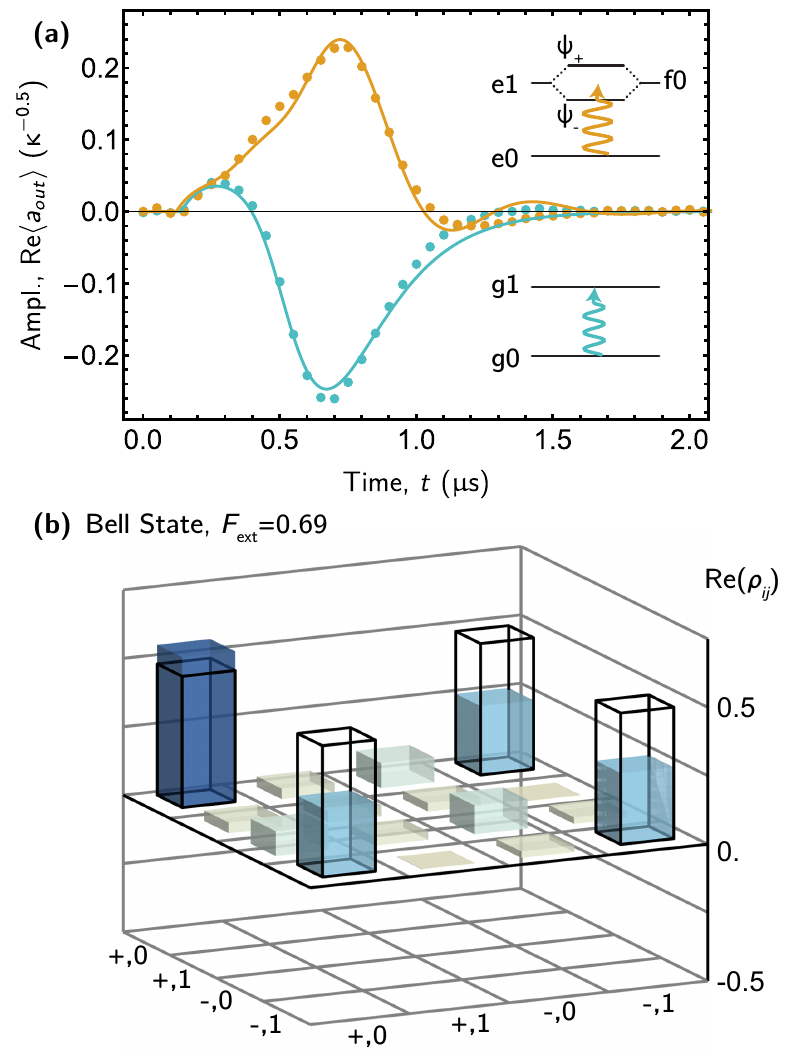}
\caption{\label{fig:cphase}
(a) Measured (dots) and simulated (lines) real part of the envelope of the amplitude $\langle a_{\text{out}} \rangle$ for the states $\ket{+}$ emitted from the source qubit and reflected off of the gate qubit while driving the $\ket{f0} \leftrightarrow \ket{e1}$ transition of the gate qubit. Level diagrams and traces are shown for the gate qubit in the ground state (turquoise) and in the excited state (orange). Imaginary part, not shown, is below $0.08/\sqrt{\kappa}$ in magnitude. (b) Real part of the measured density matrix $\rho_{\text{meas}}$ (solid bars) and the ideal density matrix $\rho_{\text{ideal}}=\ket{\psi}\bra{\psi}$ (black wireframes) of the 2-mode Bell state $\ket{\psi}=(\ket{+,0}+\ket{-,1})/\sqrt{2}$. The fidelity is $F=\bra{\psi} \rho_{\text{meas}} \ket{\psi}= 0.69$. Imaginary parts, not shown, are all below 0.04 in magnitude.
}
\end{figure}

After having confirmed our ability to absorb and re-emit photons, we demonstrate the second ingredient of the photon-photon gate scheme, shown in Fig.~\ref{fig:Principle}(c): a controlled-phase gate between the gate qubit $G$ and a photonic mode $P_i$. For this purpose, we parametrically tune the two energy levels $\ket{f0}$ and $\ket{e1}$, where $\ket{f}$ denotes the second excited state of the gate qubit, into resonance by driving the coupler with an amplitude $A$ at a frequency of $\Delta_{\text{CPHASE}}=\Delta_G+\alpha_G/2\pi=529$~MHz. Here, $\alpha_G/2\pi=302$~MHz is the anharmonicity of the gate qubit. Transitions from $\ket{e0}$ into the dressed states $\ket{\psi_+}=(\ket{f0}+\ket{e1})/\sqrt{2}$ and $\ket{\psi_-}=(\ket{f0}-\ket{e1})/\sqrt{2}$ are detuned in frequency from the bare converter mode by $g/2 \pi = 1.6$~MHz, where $g \propto A$ is the parametric interaction strength proportional to the drive amplitude $A$. For the choice of attenuators used in our setup (see App.~\ref{SM:setup}), the drive amplitude $A$ was limited by the maximal output voltage of the arbitrary waveform generator (AWG) generating the drive pulse. As in Ref.~\cite{Besse2017}, we expect an incoming photonic mode in resonance with the converter mode to be reflected off of the gate qubit with a $\pi$ phase shift when the gate qubit $G$ is in the ground state $\ket{g}$, while being reflected with no phase shift when the gate qubit $G$ is in the excited state $\ket{e}$ due to the symmetric detuning between the photonic mode and the $\ket{e0} \leftrightarrow \ket{\psi_+}$ and the $\ket{e0} \leftrightarrow \ket{\psi_-}$ transitions, see the insets in Fig.~\ref{fig:cphase}(a). We experimentally observe the predicted $\pi$ phase shift as a sign change in the real part of the average field amplitude [Fig.~\ref{fig:cphase}(a)], when preparing the gate qubit in the ground state $\ket{g}$ or in the excited state $\ket{e}$ and emitting a photonic state $\ket{+}$ from the source qubit while driving the coupler of the gate qubit at the frequency $\Delta_{\text{CPHASE}}$. The measured photon field amplitudes (dots) match well with the simulated ones, extracted from a two-level Rabi model (lines, see App.~\ref{SM:PS} for details). The simulated and measured mode profiles in Fig.~\ref{fig:cphase}(a) are, however, distorted compared to the mode profiles shown in Fig.~\ref{fig:experimental_setup}(c), since the photon bandwidth $\Gamma=\kappa_S$ is comparable to both the interaction strength $g$ and the coupling rate $\kappa_G$ of the gate qubit to the transmission line.
 
\begin{figure*}[t]
\includegraphics[width=\textwidth]{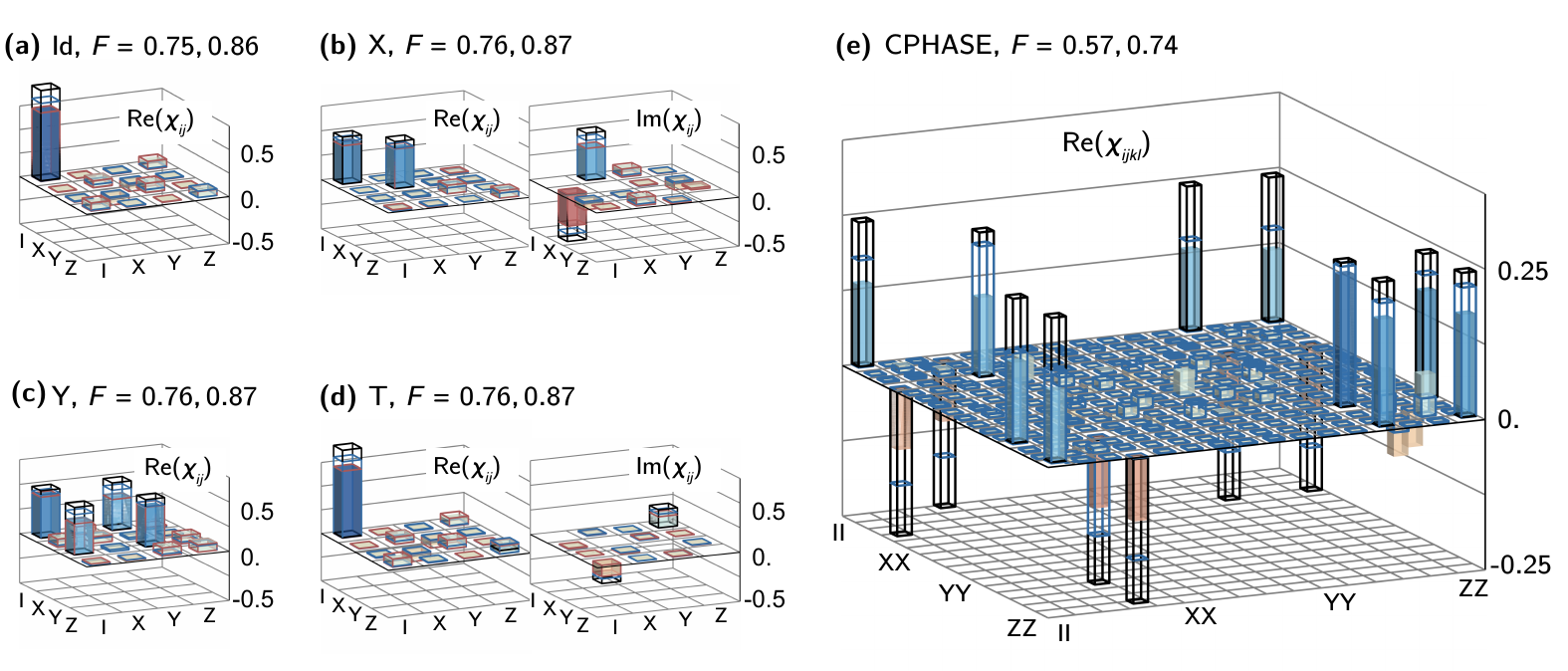}
\caption{\label{fig:qpt} Measured process maps with $\chi_{\text{int}}$ (blue wireframes) and without $\chi_{\text{meas}}$ (filled bars) accounting for state preparation errors, in comparison to simulated $\chi_{\text{sim}}$ (red wireframes) and ideal $\chi_{\text{ideal}}$ (black wireframes) process maps for (a) the identity, (b) the X-gate, (c) the Y-Gate, (d) the T-Gate and (e) the controlled-phase (CPHASE) gate. $F=F_{\text{tot}}, F_{\text{int}}$, with the total fidelity $F_{\text{tot}}=\text{Tr}(\sqrt{\sqrt{\chi_{\text{meas}}} \chi_{\text{ideal}} \sqrt{\chi_{\text{meas}}}})^2$ and the internal fidelity is $F_{\text{int}}=\text{Tr}(\sqrt{\sqrt{\chi_{\text{int}}} \chi_{\text{ideal}} \sqrt{\chi_{\text{int}}}})^2$. Imaginary parts (not shown) are all below 0.04 in magnitude.
}
\end{figure*}
Having verified all basic ingredients, we demonstrate the entangling capability of the implemented photonic controlled-phase gate [Fig.~\ref{fig:Principle}(c), see App.~\ref{SM:Calib} for detailed pulse scheme] by generating a two-mode Bell state. To do so, we perform a controlled-phase gate between the two photonic modes $P_1$ and $P_2$, each prepared in the state $\ket{+}=(\ket{0}+\ket{1})/\sqrt{2}$, measure the two field quadratures of both photonic qubits $P_1$ and $P_2$ and reconstruct the joint density matrix $\rho_{\text{meas}}$ \cite{Besse2020a,Eichler2012} of which the real parts are shown in Fig.~\ref{fig:cphase}(b). We extract a fidelity of $F=\bra{\psi} \rho_{\text{meas}} \ket{\psi}= 0.69$ to the ideal state $\ket{\psi}=(\ket{+,0}+\ket{-,1})/\sqrt{2}$, mainly limited by photon loss, mostly from the circulator, contributing $13\,\%$ to the infidelity, as well as decoherence and a finite absorption efficiency.

To further characterize the implemented single- and two-qubit gates, we perform quantum process tomography for identity, $X$-, $Y$- and $T$-gates and the CPHASE gate, which together constitute a universal set of gates \cite{Lloyd1995} acting on itinerant photonic qubits. We prepare photonic qubits in all cardinal states, perform the respective gate and reconstruct the density matrix for each input state. We then extract the process maps $\chi_{\text{meas}}$ \cite{Chuang1997}, shown in Fig.~\ref{fig:qpt} assuming ideal input states and calculate the process fidelities $F_{\text{tot}}=\text{Tr}(\sqrt{\sqrt{\chi_{\text{meas}}} \chi_{\text{ideal}} \sqrt{\chi_{\text{meas}}}})^2$ with respect to the ideal process maps $\chi_{\text{ideal}}$. We obtain close to $75\,\%$ fidelity for the single-qubit gates and $57\,\%$ for the controlled-phase gate. The obtained fidelities are small compared to state-of-the-art gate fidelities for stationary qubits directly coupled to each other on a single device \cite{Collodo2020,Barends2014,McKay2016}, mainly due to the radiation loss in the circulator and due to decoherence during the process duration, which is on the order of a $\mu$s. The performance of the photonic single- and two-qubit gates could thus be greatly enhanced by the use of less lossy circulators \cite{Lecocq2017, Chapman2017a} and by larger coupling rates.

To further analyze the obtained process maps and fidelities, we simulate the single-qubit gates using a master equation approach considering photon loss, decoherence, as well as truncation and finite bandwidth of the photonic mode (see App.~\ref{SM:MES} for details). The obtained, simulated process maps $\chi_{\text{sim}}$ have fidelities of around $97\,\%$ to the measured process maps $\chi_{\text{meas}}$ and are thus in good agreement with $\chi_{\text{meas}}$. The observed errors therefore seem largely dominated by the aforementioned mechanisms, which we account for in the master equation simulations.

Having experimentally inferred and simulated process maps, we investigate the effect of state preparation errors, specifically photon loss, on the obtained fidelities, by characterizing the internal performance of the gate set. To do so, we apply the Kraus operators $K_0= \ket{g}\bra{g}+ \sqrt{\eta_{\text{loss}}} \ket{e}\bra{e}$ and $K_1=\sqrt{1-\eta_{\text{loss}}} \ket{g}\bra{e}$ to all ideal input states, thereby generating effective input states affected by photon loss. We then extract the internal process map $\chi_{\text{int}}$ based on those effective input states and the measured density matrices (Fig.~\ref{fig:qpt}, blue wireframe). The resulting internal process fidelities $F_{\text{int}}=\text{Tr}(\sqrt{\sqrt{\chi_{\text{int}}} \chi_{\text{ideal}} \sqrt{\chi_{\text{int}}}})^2$ are close to $87\,\%$ for single-qubit gates and $74\,\%$ for the controlled-phase gate. Therefore, photon loss is the dominant error source for the presented gate set. Based on master equation simulations, we attribute most of the infidelity of the internal single-qubit processes to decoherence ($13\,\%$), while truncation of the photonic mode only contributes approximately $1\,\%$. Extending the models presented in App.~\ref{SM:MES} to also simulate process fidelities of the two-qubit gate could be an interesting subject for future theoretical work.

In summary, we implemented a universal gate set for itinerant, microwave photonic qubits. The observed internal process fidelities are close to $87\,\%$ for single-qubit gates, limited mainly by decoherence, and close to $74\,\%$ for the controlled-phase gate. The total processes, however, are still limited by photon loss. In future, this may be overcome by the development of less lossy circulators \cite{Lecocq2017, Chapman2017a} or by heralding of photons using time-bin \cite{Kurpiers2019,Lo2020a}, frequency-comb \cite{Lu2019a} or dual-rail encoding \cite{OBrien2003}. We expect the success rate of such heralding schemes to be an order of magnitude larger (close to $\eta_{\text{loss}}=75$\,\%) than in optical implementations of deterministic photon-photon gates \cite{Hacker2016b,Tiarks2019}. In addition, increasing the photon bandwidth by engineering larger coupling rates between stationary qubits and photonic qubits may reduce errors due to decoherence. We envision the demonstrated gate set to be used to generate larger entangled many-body quantum states, for example by creating additional entanglement bonds between photonic qubits \cite{Pichler2017,Wan2020b}.

\section*{Data availability statement}
The data produced in this work is available from the corresponding authors upon reasonable request.

\section*{Acknowledgments}
We thank Mihai Gabureac for contributions to the device fabrication, Baptiste Royer for input on the manuscript and the master equation simulations, and Gerhard Rempe for comments on the manuscript. This work was supported by the Swiss National Science Foundation (SNSF) through the project ``Quantum Photonics with Microwaves in Superconducting Circuits'', by the European Research Council (ERC) through the project ``Superconducting Quantum Networks'' (SuperQuNet), by the National Centre of Competence in Research ``Quantum Science and Technology'' (NCCR QSIT), a research instrument of the Swiss National Science Foundation (SNSF), and by ETH Zurich.

\section*{Author Contributions}
K.R. and J.-C.B. designed the device. J.-C.B. and G.N. fabricated the device. K.R. and J.-C.B. prepared the experimental setup. K.R. and J.-C.B characterized and calibrated the device and the experimental setup.  L.W., P.M. and P.K. implemented the photon shaping. K.R., J.-C.B. and L.W. carried out the experiments and analyzed the data, with support from P.M. and P.K.. K.R. and C.E. wrote the manuscript with input from all co-authors. A.W. and C.E. supervised the work.

\section*{Competing interests}
The authors declare no competing interests.

\begin{appendix}

%
\section{Experimental Setup and Basic Calibrations}\label{SM:setup}
\begin{figure}[!t]
\includegraphics[width=\columnwidth]{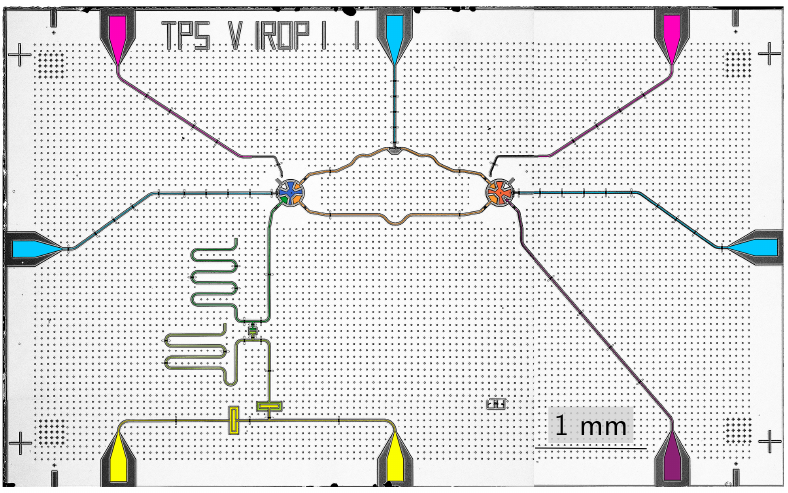}
\caption{\label{Sample}
False color optical micrograph of the sample of the gate qubit (blue). Depicted are the two coupler paths (orange) and the converter mode (dark orange), coupled to the transmission line (purple). Flux lines (cyan) and charge lines (pink) couple to the gate qubit, converter mode and the coupler. The readout resonator (green) is coupled to a feed line (yellow) via a Purcell filter (light green). 
}
\end{figure}
\begin{table}[!b]
\centering
 \begin{tabular}{l l r r}
 & & Source & Gate \\ 
 \hline
 Qubit & $g$-$e$ frequency, $\omega_{ge}/2\pi$ [GHz] & 5.925 & 5.771 \\
 & $e$-$f$ frequency, $\omega_{ef}/2\pi$ [GHz] & 5.630 &  5.478 \\
 & anharmonicity, $\alpha/2\pi$ [MHz] & 295.0 & 301.7 \\
 & lifetime of $\ket{e}$, $T_1^{(e)}$ [$\mu$s]& 16 & 13 \\
 & lifetime of $\ket{f}$, $T_1^{(f)}$ [$\mu$s]& 6 & 4 \\
 & dephasing time of $\ket{e}$, $T_2^{\star(e)}$ [$\mu$s] & 4 & 10 \\
 & dephasing time of $\ket{f}$, $T_2^{\star(f)}$ [$\mu$s] & 2 & 2 \\
 Coupler & frequency, $\omega_\text{c}/2\pi$ [GHz] & 3.2 & 4.6 \\
 Converter & $0$-$1$ frequency, $\omega_{01}/2\pi$ [GHz] & 5.998 & 5.998 \\
 & decay rate, $\kappa/2\pi$ [MHz] & 1.8 & 2.1 \\
 \hline
 \end{tabular}
 \caption{\label{table:paramp}Measured device parameters.}
\end{table}
\begin{figure*}[!t]
\includegraphics[width=\textwidth]{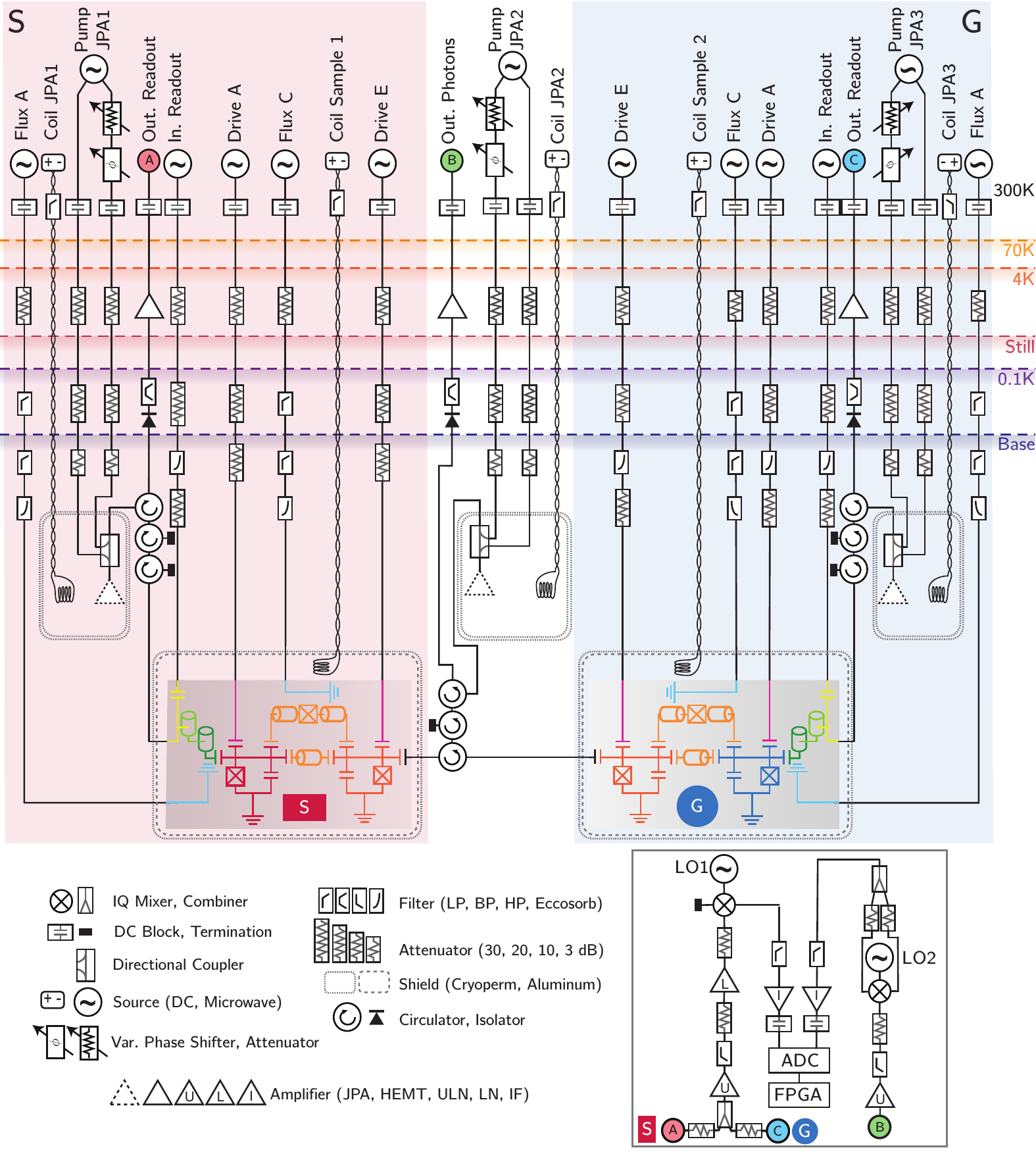}
\caption{\label{fig:Setup}
Experimental setup for operating the source qubit (S, red) and the gate qubit (G, blue) devices. For details see text.
}
\end{figure*}
We fabricated two 4.3~mm x 7~mm samples, both based on the same base layer design, one for the source qubit and one for the gate qubit. The sample for the gate qubit was also used in Ref.~\cite{Besse2020a}, see Fig.~\ref{Sample} for a false color optical micrograph. Using optical photolithography and reactive ion etching, we pattern qubits pads and coplanar waveguide structures into a niobium film sputtered on a silicon substrate. To fabricate the Al/AlOx/Al Josephson junctions, we use electron-beam lithography and shadow evaporation. We mount the packaged samples onto the base temperature stage (20$\,$mK) of a dilution refrigerator, each housed inside its dedicated set of magnetic shields made of aluminum and cryoperm, see sketch of the experimental setup in Fig.~\ref{fig:Setup}. The two samples are connected via coaxial cables and a circulator, which is also thermalized to the base plate. We apply microwave pulses to the source and gate qubit and to the respective readout resonators via charge lines incorporating 20~dB of attenuation at the 4 K, 100 mK and base temperature stage for signal conditioning \cite{Krinner2019}. We use phase-preserving Josephson Parametric Amplifiers (JPAs) \cite{Eichler2014a} to amplify both the gate-qubit-readout signal and the field associated with the photonic qubits with a gain of 25~dB and 14~dB, respectively. The JPA for the source-qubit readout was not used. All JPAs are mounted in a pair of cryoperm shields. We further amplify the detection signals using a high-electron mobility transistor (HEMT) at the 4~K stage, and room-temperature amplifiers. Compared to the readout lines, the photon detection line has one amplification stage less to preserve the linearity in the amplification chain, which is a crucial requirement for photon field tomography~\cite{Eichler2012}.

\begin{figure}[!t]
\includegraphics[width=\columnwidth]{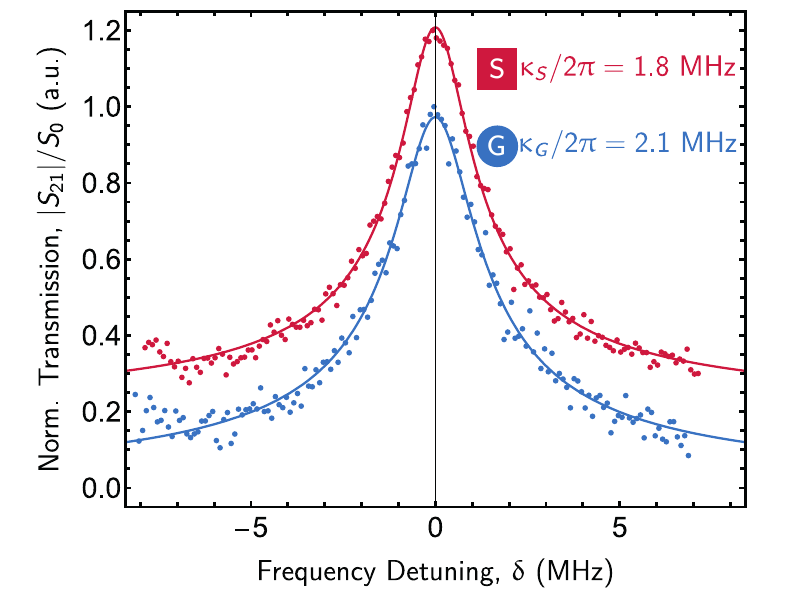}
\caption{\label{fig:Kappa}
Absolute value of the measured transmission coefficient $|S_{21}|/S_0$, normalized to the transmission at zero detuning $S_0$, when driving the converter mode of the gate qubit (blue) and the source qubit (red, vertically offset by 0.2) via their respective charge lines. Solid lines are Lorentzian fits to the data, see text, resulting in the indicated linewidths $\kappa$. 
}
\end{figure}

\begin{figure}[b]\setlength{\hfuzz}{1.1\columnwidth}
\begin{minipage}{\textwidth}
\includegraphics[width=\textwidth]{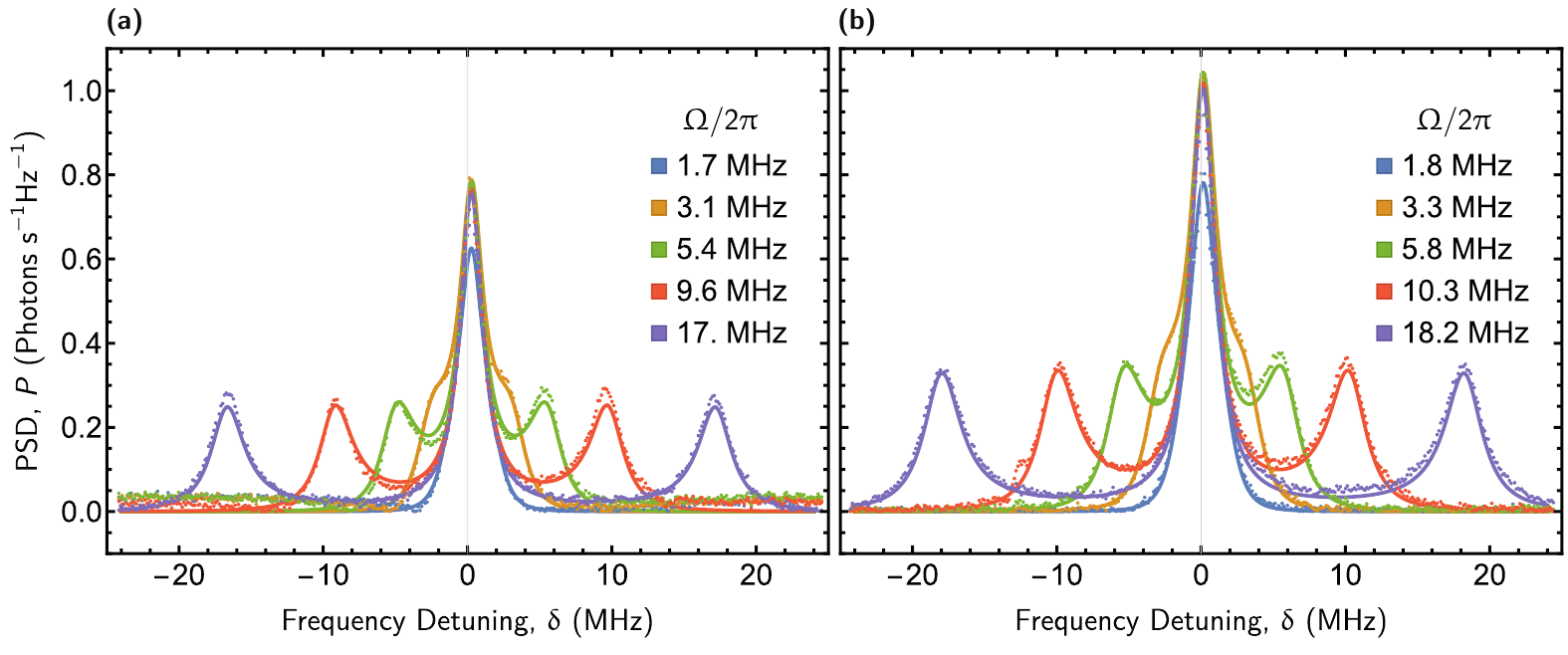}
\caption{\label{fig:MollowFig}
Measured power spectral density (PSD) $P$ of the inelastic scattering of a coherent tone resonant with (a) the source qubit and (b) the gate qubit converter mode for the indicated drive rates $\Omega$. All data sets are scaled with respect to the gate qubit fitted amplitude $P_{0,G}$, see text. Solid lines are joined fits to all data sets with a single set of fitting parameters.
}
\end{minipage}
\end{figure}

We measure the basic device parameters presented in Table~\ref{table:paramp}, by performing one- and two-tone spectroscopy measurements, as well as Rabi, Ramsey and coherence measurements, as presented in Ref.~\cite{Besse2020a}. Both the gate and the source qubit are tuned to their respective maximum frequency \cite{Koch2007}. We bias both couplers to the point, at which the effective coupling between the source/gate qubit and the converter mode $J$ vanishes. The converter modes are both tuned to $\omega_{01}/2\pi =5.998$~GHz. We extract the decay rate $\kappa$ for the respective converter modes by measuring the coherent scattering of a weak input tone into the transmission line \emph{vs.}~frequency and by fitting the data to the function $|S_{21}|=|S_0/(1+2 i \delta 2 \pi / \kappa)|$, obtained from input-output theory with $\delta$ being the frequency detuning between the converter mode frequency $\omega_{01}/2\pi$ and the probe frequency and $S_0$ being the transmission at zero detuning, see Fig.~\ref{fig:Kappa}.


\begin{figure*}[!t]
\includegraphics[width=\textwidth]{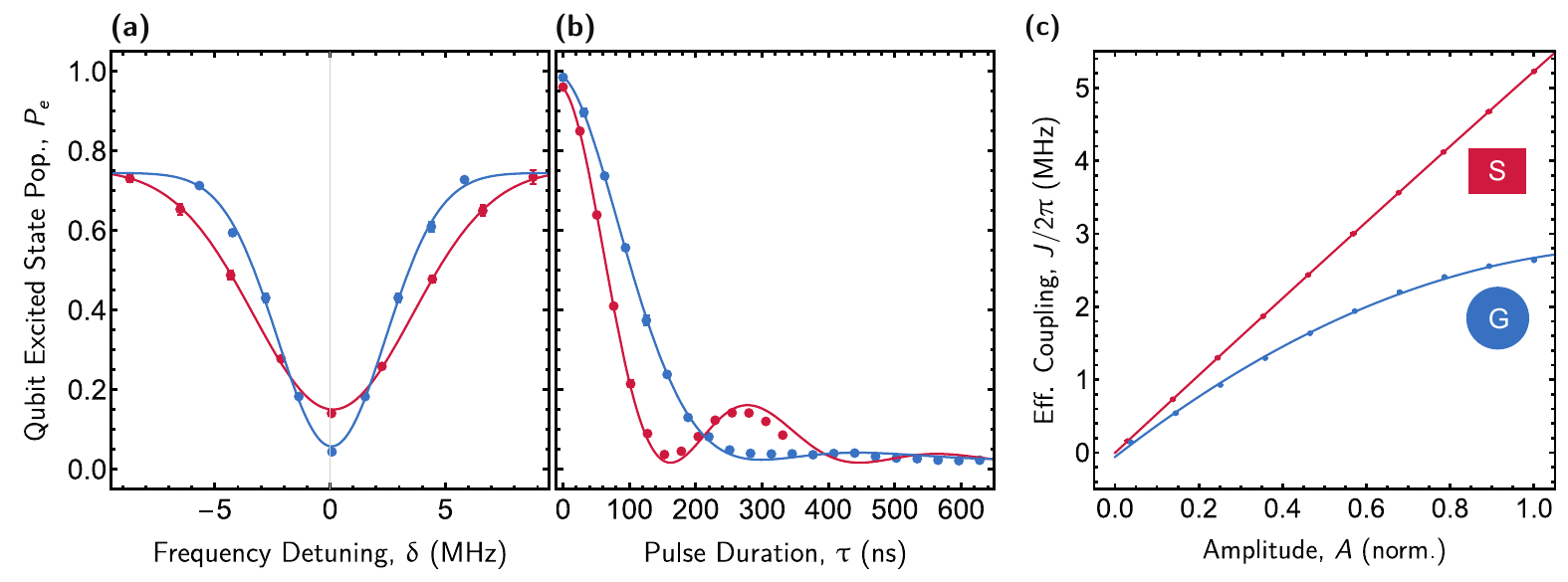}
\caption{\label{fig:Rabi}
(a,b) Excited state population $P_e$ after preparing the source qubit (red) [gate qubit (blue)] in the excited state and applying a modulated flux pulse with a square envelope and a fixed, normalized amplitude of $A=0.35$. (a) Fixed pulse duration $\tau=283$~ns ($349$~ns), plotted \emph{vs.}~flux pulse frequency detuning $\delta$ from the $\ket{e0} \leftrightarrow \ket{g1}$ transition. Solid lines are Gaussian fits to the data. (b) On resonance $\delta=0$, plotted \emph{vs.}~flux pulse duration $\tau$. Solid lines are fits to the data, see text. (c) Effective coupling rate $J$ for the $\ket{e0} \leftrightarrow \ket{g1}$ transition of the gate qubit (G, blue circle) and the source qubit (S, red rectangle) \emph{vs.}~drive amplitudes $A$, normalized to the respective maximal amplitude. Solid lines are quadratic fits to data. 
}
\end{figure*}

To calibrate the power of the emitted photonic qubits, we measure the incoherent scattering while driving the converter mode resonantly at the set of indicated drive powers, observing Mollow triplets, see Fig.~\ref{fig:MollowFig}. We globally fit the observed data to the expected power spectral density function $P$ \emph{vs.}~the detuning from the drive frequency $\delta$, obtained by considering a dressed two-level system \cite{Mollow1969,Lang2011}:
\begin{align}
P(\delta)&= P_0 \kappa \frac{64 \kappa \Omega^4 (2 \kappa^2 + 2(\delta-f_0)^2+\Omega^2)}{\pi (\kappa^2 +4(\delta-f_0)^2)(\kappa^2+2 \Omega^2) p_+ p_-}, \text{ with} \\
p_{\pm}&=(5 \kappa^2 + 8(\delta-f_0)^2 - 8 \Omega^2 \pm 3 \kappa \sqrt{\kappa^2-16 \Omega^2}), \notag
\end{align}
where $\kappa=\kappa_{S/G}$ is the decay rate of the respective converter mode, $\Omega$ is the drive rate and $f_0$ takes into account small detunings between the drive frequency and the converter mode. From the fitted amplitudes $P_0$ for the source device $P_{0,S}$ and the gate device $P_{0,G}$, we extract the transmission efficiency expected from radiation loss between the source and the gate chip of $\eta_{\text{loss}}=P_{0,S}/P_{0,G}=75\,\%$.
\unskip\parfillskip 0pt \par
\newpage
\noindent

\section{Photon Shaping}\label{SM:PS}
\subsection{Concept}
To optimize the absorption efficiency, we generate photonic fields with a near time-symmetric temporal profile by controlling the time evolution of the tunable, effective coupling rate $J$ between the qubit and its converter mode, following the protocol detailed in Refs.~\cite{Kurpiers2018,Pechal2014,Morin2019}. We choose the temporal envelope
\begin{align}
\xi(t)&= \frac{\sqrt{\Gamma}}{2} \text{sech}(\Gamma t/2),
\label{shape}
\end{align}
where the effective bandwidth $\Gamma$ is bound by the coupling rate $\kappa$ of the converter mode: $\Gamma \leq \kappa$. To emit photonic fields with the temporal envelope $\xi(t)$, the effective coupling rate is controlled according to \cite{Kurpiers2018}:
\begin{align}
J(t)&=\frac{\Gamma (-e^{\Gamma t}+1+\kappa (1+e^{\Gamma t})/\Gamma )}{4  \cdot \text{cosh}(\Gamma t/2)\sqrt{(1+e^{\Gamma t})\kappa/\Gamma-e^{\Gamma t}}}.  \label{js}
\end{align}
To controllably absorb photonic fields with an envelope $\xi(t)$, we apply a time-reversed coupling $J(-t)$.

We choose to emit shaped photons with the maximal effective bandwidth $\Gamma$, thereby minimizing errors due to decoherence. Since we emit photonic qubits from the source qubit $S$ and absorb them at the gate qubit $G$, $\Gamma$ is limited by the smaller coupling rate of the two converter modes $\Gamma \leq \min{[\kappa_S,\kappa_G]}$. In our setup, $\kappa_S<\kappa_G$. Therefore, the effective photon bandwidth $\Gamma$ is limited by the coupling rate $\kappa_S$ of the source qubit converter mode.

\subsection{Implementation}
We control the time evolution of the effective coupling rate $J$ by applying an alternating current (AC) pulse $I(t)$ to the flux line of the respective coupler. We generate the AC pulse $I(t)$ using an arbitrary waveform generator (AWG). We combine the AC pulse with the direct current (DC) offset applied for biasing the coupler and route the combined pulse into the coupler flux line. The AC pulse $I(t)$ induces an alternating magnetic flux $\Phi(t)$ in the SQUID loop of the coupler. As a result, the coupler frequency oscillates at the frequency of the AC pulse $I(t)$. When the AC pulse $I(t)$ is in resonance with the $\ket{e0} \leftrightarrow \ket{g1}$ transition frequency, we control the amplitude and the phase of the effective coupling rate $J(t)$.

\begin{figure}[!t]
\includegraphics[width=\columnwidth]{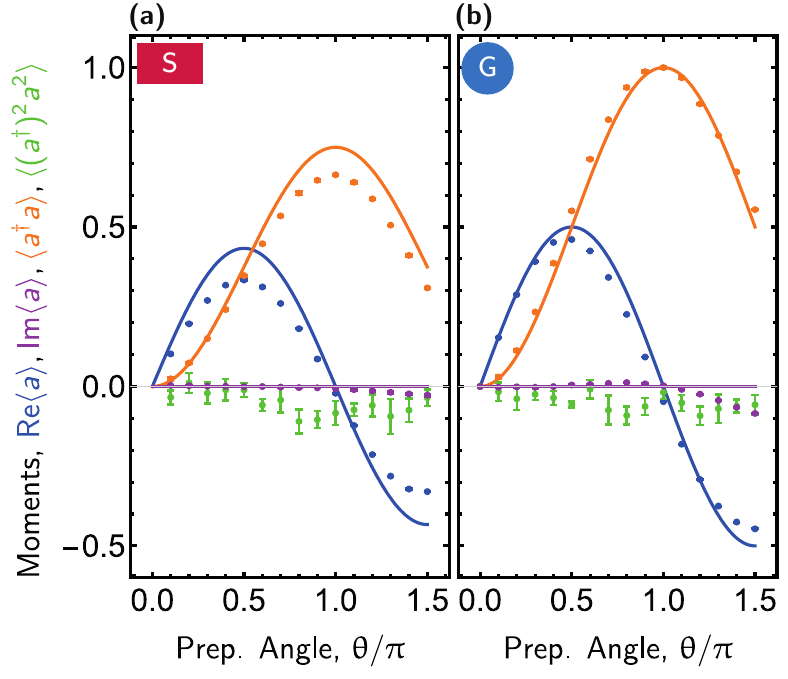}
\caption{\label{fig:moments} Measured moments $\text{Re}\langle a\rangle$ (blue), $\text{Im}\langle a\rangle$ (purple), $\langle a^\dagger a \rangle$ (orange), and $\langle (a^\dagger)^2 a^2 \rangle$ (green), for superpositions $\ket{\psi}=\sin (\theta/2)\ket{0}+\cos (\theta/2)\ket{1}$ of vacuum and single-photon states emitted from (a) the source qubit and (b) the gate qubit. Lines indicate ideal expectation values, calculated taking into account photon loss between the source and the gate chip.
}
\end{figure}

\begin{figure}[b]\setlength{\hfuzz}{1.1\columnwidth}
\begin{minipage}{\textwidth}
\includegraphics[width=\textwidth]{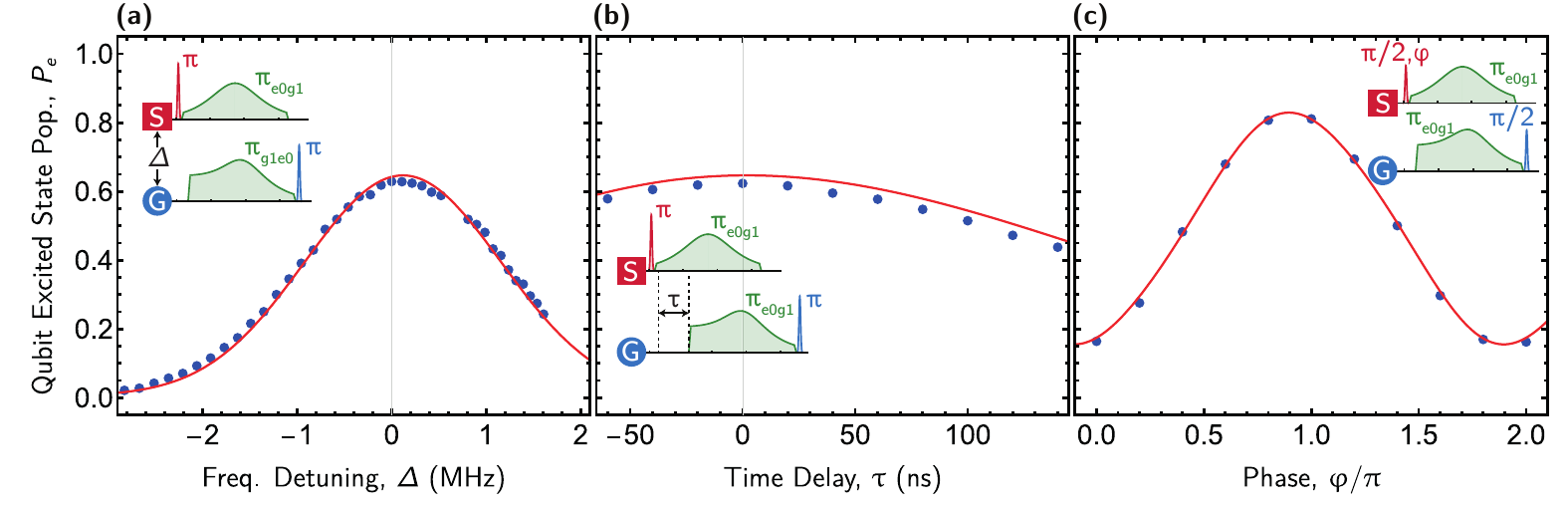}
\caption{\label{fig:PhotonGatesCalib}
Measured (blue dots) and simulated (red line) excited state population $P_e$ of the gate qubit for three calibration routines. The inset shows the envelopes of the pulse sequence applied to the gate chip (G, blue circle) and the source chip (S, red rectangle) during the respective calibration steps. (a) Measured excited state population $P_e$ after emitting a single-photon Fock state $\ket{1}$ from the source qubit and absorbing it at the gate qubit \emph{vs.}~frequency detuning $\Delta$ between the converter modes of the source and gate qubit. Photons are emitted/absorbed in resonance with the respective converter mode. (b) Excited state population $P_e$ \emph{vs.}~time delay $\tau$ (relative to the optimal time delay found in the experiment) between the emission pulse at the source qubit and the absorption pulse at the gate qubit. (c) Excited state population $P_e$, when emitting and absorbing a photonic superposition state $(\ket{0}+ e^{i \varphi }\ket{1})/\sqrt{2}$ \emph{vs.}~phase $\varphi$ and applying a final $\pi/2$ rotation along the Y axis to the gate qubit.
}
\end{minipage}
\end{figure}

To tune the AC pulse into resonance with the $\ket{e0} \leftrightarrow \ket{g1}$ transition frequency, we prepare the qubit in the excited state and measure the population remaining after applying a modulated square pulse  $I_{\square}(t)=A \cos(\omega t) [\Theta(0)-\Theta(\tau)]$ with varying frequency $\omega/2 \pi$, see Fig.~\ref{fig:Rabi}(a), for a range of pulse amplitudes $A$ at a fixed pulse duration $\tau$. $\Theta(t)$ indicates the Heaviside step function. We determine the frequency of the $\ket{e0} \leftrightarrow \ket{g1}$ transition as the one for which the excited state population is minimized. Since we drive an excitation-number conserving transition via a tunable coupler, we avoid AC Stark effects as compared to the scheme in Refs.~\cite{Pechal2014, Kurpiers2018}.

To measure the effective coupling rate $J$, we vary the duration $\tau$ of the square pulse $I_{\square}(t)$ at fixed, resonant frequency $\omega/2 \pi$, see Fig.~\ref{fig:Rabi}(b) for $A=0.35$. By considering a driven two-level system decaying into the ground state with rate $\kappa$ \cite{Kurpiers2018}, we fit the measured population \emph{vs.}~duration $\tau$ to obtain the effective coupling rate $J$. From the same fit, we also extract the coupling rates of the converter modes to the transmission line $\kappa_S/2\pi=2.1$~MHz and $\kappa_G/2\pi=2.8$~MHz, which deviate from the coupling rates obtained by spectroscopy (Fig.~\ref{fig:Kappa}). The nature of this deviation is yet to be investigated. We used the fitted coupling rates $\kappa_S/2\pi=2.1$~MHz and  $\kappa_G/2\pi=2.8$~MHz for the generation of coupler pulses $I_{\xi}(t)$. Finally, we obtain a mapping between the effective coupling rate $J$ and the amplitude of the square pulse $A$ by quadratically interpolating the fitted coupling rates \emph{vs.}~amplitude $A$, shown in Fig.~\ref{fig:Rabi}(c). By inverting the mapping $A \rightarrow J$, we convert the targeted time evolution of the coupling rate $J(t)$ in Eq.~(\ref{js}) into a coupler pulse $I_{\xi}(t)$, that is generated by the AWG.

To minimize the effects of decoherence, we truncate the coupler pulse $I_{\xi}(t)$ at $t=\pm 4.6 / \Gamma$. We estimate that the photon shape generated by the truncated pulse has an overlap of $98\,\%$ with the targeted envelope. For the estimate, we numerically evaluate a two-level Rabi model decaying into the ground state with rate $\kappa$, while being driven by the coupler pulse $I_{\xi}(t)$. The overlap provides a good estimate of the absorption efficiency \cite{Kurpiers2018, Morin2019}. The two-level Rabi model was also used to extract the expected temporal profiles shown as solid lines in Fig.~\ref{fig:experimental_setup}(c). To extract the expected temporal profiles after reflection of a photonic mode off the gate qubit [shown in Fig.~\ref{fig:cphase}(a)], we calculate the reflection coefficient $S_{11}$ \emph{vs.}~frequency of the gate device while driving the $\ket{f0} \leftrightarrow \ket{e1}$ transition, for the gate qubit in the ground state $\ket{g}$ and excited state $\ket{e}$ using input-output theory. We then convolve the temporal profile of the incoming photonic mode with the response function for the gate qubit in the ground state $\ket{g}$ and excited state $\ket{e}$, obtained by Fourier-transforming the respective spectrum $S_{11}$.

\subsection{Single-Photon Tomography}
To verify the single-photon character of the photonic qubits, we prepare the source and the gate qubit in the state $\sin (\theta/2)\ket{0}+\cos (\theta/2)\ket{1}$ for the indicated range
\unskip\parfillskip 0pt \par
\newpage
\noindent of angles $\theta$ and apply a swap gate using the coupler pulse $I_{\xi}(t)$. We measure the generated photonic qubits using heterodyne detection and extract moments up to 4th order \cite{Eichler2012}, shown in Fig.~\ref{fig:moments}, normalized by the value of $\langle a^\dagger a \rangle$ for $\theta=\pi$ for the gate qubit. For both qubits, we find that $\langle (a^\dagger)^2 a^2 \rangle$ is close to zero, but slightly negative. While we have not studied the origin of this effect in detail, a weak gain compression in the amplification chain on the order of $0.01$~dB caused by the single-photon signal at its input could explain the observed negative values in the inferred $\langle (a^\dagger)^2 a^2 \rangle$ correlations. The amplitude $\text{Re}\langle a\rangle$ is reduced, compared to the ideal value, due to dephasing during the emission process. The source qubit has a smaller dephasing time ($T_{2,S}^{\star(e)}=4$~$\mu$s) than the gate qubit ($T_{2,G}^{\star(e)}=10$~$\mu$s), thus the reduction in the amplitude $\text{Re}\langle a\rangle$ is more pronounced. Error bars were obtained by evaluating five independent measurements. Moments $\langle (a^\dagger)^n a^m \rangle$ for the source qubit are reduced by a factor of $\eta_{\text{loss}}^{\frac{n+m}{2}}=0.75^{\frac{n+m}{2}}$ due to photon loss between the source and the gate qubit.


\section{Photon Gate Calibration and Pulse Scheme}\label{SM:Calib}
Having calibrated the emission process for both source and gate qubit individually, we optimize the simultaneous emission and absorption process to maximize the absorption efficiency. Specifically, we emit a single-photon Fock state $\ket{1}$ from the source qubit, absorb it at the gate qubit, and read out the state of the gate qubit. We repeat this transfer scheme while sweeping the experimental parameters detailed hereafter.

\begin{figure}[!t]
\includegraphics[width=\columnwidth]{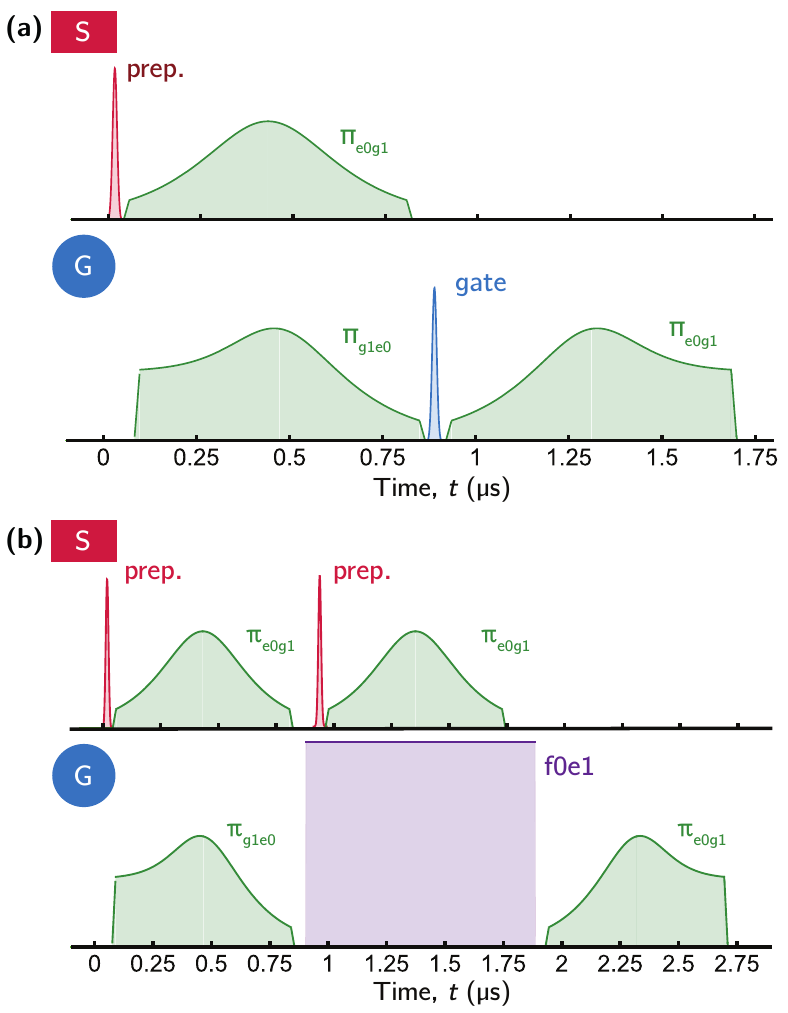}
\caption{\label{fig:Pulses} Timing diagram of the pulses applied to the gate qubit (G, blue circle) and the source qubit (S, red rectangle) for quantum process tomography of (a) single-photon gates and (b) the CPHASE gate. Depicted are the envelopes of the pulses on the charge line of the source (red) and the gate qubit (blue), as well as the envelopes of the flux pulses driving the $\ket{e0} \leftrightarrow \ket{g1}$ transition with the flux line of the respective coupler (green) and the $\ket{f0} \leftrightarrow \ket{e1}$ transition via the flux line of the coupler of the gate qubit (purple).
}
\end{figure}

\begin{figure}[!t]
\includegraphics[width=\columnwidth]{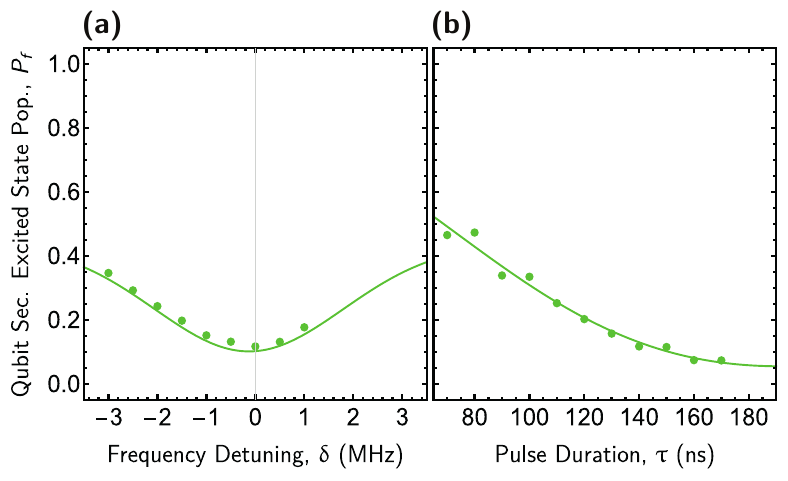}
\caption{\label{fig:ChevronF} (a,b) Second excited state population $P_f$ after preparing the gate qubit in the second excited state and applying a modulated flux pulse with a square envelope and a fixed (normalized) amplitude of $A=1$. (a) Fixed pulse duration $\tau=170$~ns, plotted \emph{vs.}~flux pulse frequency detuning $\delta$ from the $\ket{f0} \leftrightarrow \ket{e1}$ transition. Solid line is a Gaussian fit to the data. (b) On resonance $\delta=0$, plotted \emph{vs.}~flux pulse duration $\tau$. Solid line is a fit to the data based on a driven two-level system decaying at a rate $\kappa_G/2\pi=2.1$~MHz.
}
\end{figure}

First, to ensure that both converter modes are exactly in resonance, we maximize the excited state population at the gate qubit, while sweeping the gate qubit converter mode frequency, see Fig.~\ref{fig:PhotonGatesCalib}(a). For each converter mode frequency, we repeat the calibrations for emitting/absorbing shaped photons, as detailed in App.~$\ref{SM:PS}$. Second, we calibrate the timing of pulses by sweeping the delay between emission and absorption pulses, shown in Fig.~\ref{fig:PhotonGatesCalib}(b). For the optimal time delay, the excited state population is again maximal and sets the optimized transfer efficiency of $64\,\%$. Third, we calibrate the unknown phase offset, originating from the optical path length between the two chips. To do so, we emit and absorb a superposition state $(\ket{0}+ e^{i \varphi }\ket{1})/\sqrt{2}$ sweeping the phase $\varphi$, apply a $\pi/2$ pulse to the gate qubit and extract the excited state populations at the gate qubit $G$, shown in Fig.~\ref{fig:PhotonGatesCalib}(c). Master equation simulations (see App.~\ref{SM:MES}) are in good agreement with the measured calibration datasets.

Subsequently, we implement the universal gate set, consisting of the identity, the $X-$, the $Y-$, the $T-$ and the controlled-phase gate in our setup using the pulse schemes shown in Fig.~\ref{fig:Pulses}(a) for single-photon gates and Fig.~\ref{fig:Pulses}(b) for the photon-photon gate. Coupler pulses $I_{\xi}(t)$ for emission and absorption (green in Fig.~\ref{fig:Pulses}) are applied to the respective coupler. We realize $X-$ and $Y-$ gates on the transmon qubits by using Derivative Removal by Adiabatic Gate (DRAG) pulses \cite{Motzoi2009} with amplitude $\theta=\pi/2$ and a phase of $\varphi=\pi/2$ and $\varphi=0$ respectively, truncated at $\pm3\sigma$ where $\sigma=6$~ns is the linewidth of the pulse. The identity gate is realized by waiting $t=36$~ns, such that the time bin of the photonic qubits is independent of the applied gate. The $T-$gate is realized as a virtual $T-$gate after waiting $t=36$~ns by changing the reference frame for the subsequent coupler pulse $I_{\xi}(t)$, effectively adding a phase of $\pi/8$ to $I_{\xi}(t)$.

During the controlled-phase gate, we apply a modulated square pulse with maximal amplitude (limited by the dynamic range of the arbitrary waveform generator) to drive the $\ket{f0}\leftrightarrow \ket{e1}$ transition. To calibrate the controlled-phase gate, we sweep frequency and pulse duration of the modulated square pulse, and readout the remaining populations of the gate qubit, see Fig.~\ref{fig:ChevronF}. From the measured populations in Fig.~\ref{fig:ChevronF}(b), we extract an interaction strength $g/2\pi=1.6$~MHz.

\vspace{20mm}
\section{Master Equation Simulations}\label{SM:MES}
To simulate single-qubit process maps and calibration routines, we use a master equation approach, extending on the work done in Ref.~\cite{Kurpiers2018}. For all master equation simulations, we use the python package QuTiP \cite{Johansson2013a}.

To simplify the simulation, we model both the gate and the source qubit as anharmonic resonators with three energy levels and anharmonicity $\alpha$ coupled to a converter mode with two energy levels. In a frame rotating at both the $\ket{g}\leftrightarrow \ket{e}$ frequency $\omega_{ge}/2\pi$ of the qubit and the $\ket{0}\leftrightarrow \ket{1}$ frequency $\omega_{01}/2\pi$ of the respective converter mode, driving the $\ket{e0} \leftrightarrow \ket{g1}$ transition can be modeled by a resonant coupling term. We thus arrive at the Hamiltonian $H$
\begin{align}
H&=\sum_{i=S,G} \left[ \frac{\alpha_i}{2} a_i^{\dagger} a_i^{\dagger} a_i a_i \right] \label{anharm} \\
&- \frac{i}{2} \kappa_S \kappa_G \eta_{\text{loss}} (b_S b_G^{\dagger} - b_G b_S^{\dagger}) \label{direct}  \\
&+ J_S(t) (a_S b_S^{\dagger}+a_S^{\dagger} b_S) \label{JS}\\
&+ J_G(t) (a_G b_G^{\dagger}+a_G^{\dagger} b_G), \label{JD}
\end{align}
where the indices denote the source qubit $S$ and the gate qubit $G$, $a$ denotes the annihilation operator of the respective qubit, $b$ denotes annihilation operator of the converter mode, $\kappa$ denotes the coupling rates of the converter modes to the transmission line and $\eta_{\text{loss}}=0.75$ is the photon transmission probability.

The first term (\ref{anharm}) in the Hamiltonian $H$ accounts for the detuned second excited state $\ket{f}$. We introduce the time-dependent coupling terms (\ref{JS}) and (\ref{JD}) to model the parametric coupling via the $\ket{e0} \leftrightarrow \ket{g1}$ transition. The second term (\ref{direct}) of the Hamiltonian, together with the collapse operators \cite{Kurpiers2018,Kiilerich2019}
\begin{align}
c_1&=\sqrt{\kappa_S \eta_{\text{loss}}} b_S+\sqrt{\kappa_G} b_G \label{c1}\\
c_2&=\sqrt{\kappa_S} (1-\eta_{\text{loss}}) b_S, \label{c2}
\end{align}
introduce a directional coupling between the two converter modes \cite{Carmichael1993} and thus model a circulator with $1-\eta_{\text{loss}}=25$\,\% loss. We take into account decoherence by including collapse operators for dephasing and relaxation in our simulation. For the simulation, we use the measured device parameters as specified in Table~\ref{table:paramp}. We evolve the time-dependent coupling terms $J(t)$ in the simulations according to Eq.~(\ref{js}), taking into account the truncation and parameters for the coupler pulses $I_{\xi}(t)$ used in the experiment.

In this way, we extract both the simulated process maps $\chi_{\text{sim}}$, shown in Fig.~\ref{fig:qpt}, and the simulated excited state populations for the calibration routines shown in Fig.~\ref{fig:PhotonGatesCalib}. We do not account for imperfections in the emission process from the gate qubit, since a photon Fock state $\ket{1}$ emitted from the gate qubit serves as a reference for the reconstruction of the measured density matrices \cite{Besse2020a,Eichler2012}.



\end{appendix}



\bibliography{PhotonGates}

\end{document}